# Computational Analysis of a Spiral Thermoelectric Nanoantenna for Solar Energy Harvesting Applications

Edgar Briones[1], Joel Briones[2], Alexander Cuadrado[3], Stefan McMurtry[4], Michel Hehn[4], François Montaigne [4], Javier Alda[3], Francisco Javier González[1]

[1] Coordinación para la Innovación y Aplicación de la Ciencia y la Tecnología, Universidad Autónoma de San Luís Potosí, SLP, México, edgarbriones@yahoo.com
[2] Departamento de Fisica, Universidad de Santiago de Chile (USACH), 9170124 Santiago, Chile
[3] Optics Complutense Group, University Complutense of Madrid, Faculty of Optics and Optometry, Madrid. Spain
[4] Institut Jean Lamour, CNRS, Université de Lorraine, Bd des Aiguillettes, BP70239, F-54506 Vandoeuvre Les Nancy, France

*Abstract*—Thermo-electrical nanoantennas have been proposed as an alternative option for conversion solar energy harvesting applications. In this work, the response of a spiral broadband antenna has been obtained from numerical and theoretical simulations perspectives. The results show that this device exhibits a responsivity of 20mV/W under 117W/cm$^2$, for a single-frequency radiation. We discuss strategies for enhanced efficiency.

*Index Terms*— nanoantennas, Seebeck effect, solar energy harvesting.

## I. Introduction

Nano-antennas are resonant metallic structures that confine the optical energy into small volumes in an efficient way by inducing a high-frequency current in its structure [1-3]. Due to the high light-absorbing efficiencies reached by these structures, nano-antennas have gained considerable attention over the last years for the engineering of novel photovoltaic devices at visible and infrared wavelengths [4-8].

However, a successful implementation of these elements for photovoltaic applications has not been achieved yet due to the lack of an efficient harvesting mechanism of the optical energy. In this context, nanoantennas coupled to high-speed rectifiers based on tunnel barriers, also known as "rectennas", has been extensively explored during the last years as a way to convert the optical power into dc power [9-13], nonetheless, these devices exhibit low efficiencies due to the poor performance of those rectifiers at optical frequencies [14].

In this work, thermoelectric Ti-Ni nanoantennas are proposed as an alternative strategy to recover the optical energy coupled to them. These devices are based on the very basic principle of a thermocouple operating by the Seebeck effect, where a voltage drop is induced all along the structure by locally increasing the temperature of the bi-metallic junction [15]. In must be pointed out that here, the heating of the junction is optically induced when the optical radiation impinges the structure, causing a non-uniform Joule heating generated by the resonant current along the nanoantennas [16, 17]. In this work, the responsivity and the optical-to-electrical conversion efficiency of these types of devices is evaluated numerically by using COMSOL Multi-Physics.

## II. Method

A sketch of the proposed antennas structure is shown in Fig. 1. We choose a broadband square spiral-antenna tuned to absorb at thermal-infrared wavelengths, designed using two symmetrical arms composed by seven elements of 200nm wide and 200nm thick, as defined in [3]. The arms of the spiral antenna were built of titanium and nickel, metallic materials that combine both a low thermal conductivity with considerably thermo-power properties [18]. The structure was placed on a semi-infinite $SiO_2$ substrate and a 10.6μm wavelength plane-wave was used for far-field illumination (of arbitrary irradiance W=117W/cm$^2$), whose polarization matched the circular polarization of the square spiral nanoantenna.

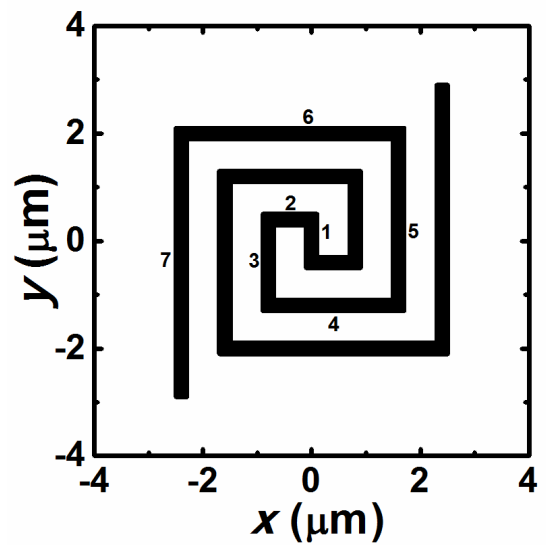

Fig. 1. Geometry of a two arms square spiral antenna tuned to absorb the thermal infrared wavelengths at 10.6μm.

The numerical analysis was performed by using a multi-physics approach where both the electromagnetic and thermal domains are fully integrated. The numerical model was built using all the optical, thermal and thermo-electrical properties of materials reported at 10.6μm wavelength [18, 19].

As a first step the optically-induced heat along the structure was evaluated by considering the computation as a full-optical problem, since heat is generated by Joule effect. For this, the power density $q(r)$ was evaluated inside the volume of the nanoantennas by [20]

$$q(r)_{antenna} = (1/2)\, \sigma(\omega)\, |E(r)|^2 \quad (1)$$

where $\sigma(\omega)$ is the conductivity of the material at the frequency of the incident wave and $E(r)$ the distribution of the electric field inside the nano-antennas when illuminated.

The power density was then used as a source of heat into the solver in order to perform numerical simulations of the steady-state temperature distribution T inside and outside the resonant metallic structures. The temperature changes are calculated by solving the heat transfer equation

$$\nabla \cdot [\kappa \nabla T(r)] = -q(r) \quad (2)$$

by taking $q(r)$ as $q(r)_{antenna}$ and 0, inside and outside the nanostructures, respectively. The effective temperature increase of the bi-metallic junction ΔT was found and the voltage response of the antenna was evaluated by using the Seebeck effect relationship [17]

$$V_{OC} = (S_{Ti} - S_{Ni})\, \Delta T \quad (3)$$

where $S_{Ti}$ and $S_{Ni}$ are the Seebeck coefficients for the nickel and titanium taken at bulk ($S_{Ti}$=7.2μV/K and $S_{Ni}$=-15μV/K). Finally, the responsivity exhibited by the thermo-electric nano-antennas was evaluated by

$$\Re_V = V_{OC}/P_{inc} \quad (4)$$

where $P_{inc}$ is the power contained by the incident wave.

In spite that the thermal-optical analysis is performed by a classic treatment (neither thermal nor electric contribution arising from interfaces are taken into account), this method has shown to give accuracy results for thermal-optical analysis of nanoantennas coupled-bolometers [21, 22].

### III. RESULTS

The distribution temperature of the Ti-Ni square spiral nano-antenna at the steady-state is shown in Fig. 2 (a) and (b). As expected, an important locally heating of the bi-metallic interface is achieved due to the strong built electric resonant current exhibited by the spiral nano-antennas at the mid-point. The temperature asymmetry of arms arises from the difference in thermal conductivities of used metals what causes one arm to conduct faster the heat to the substrate.

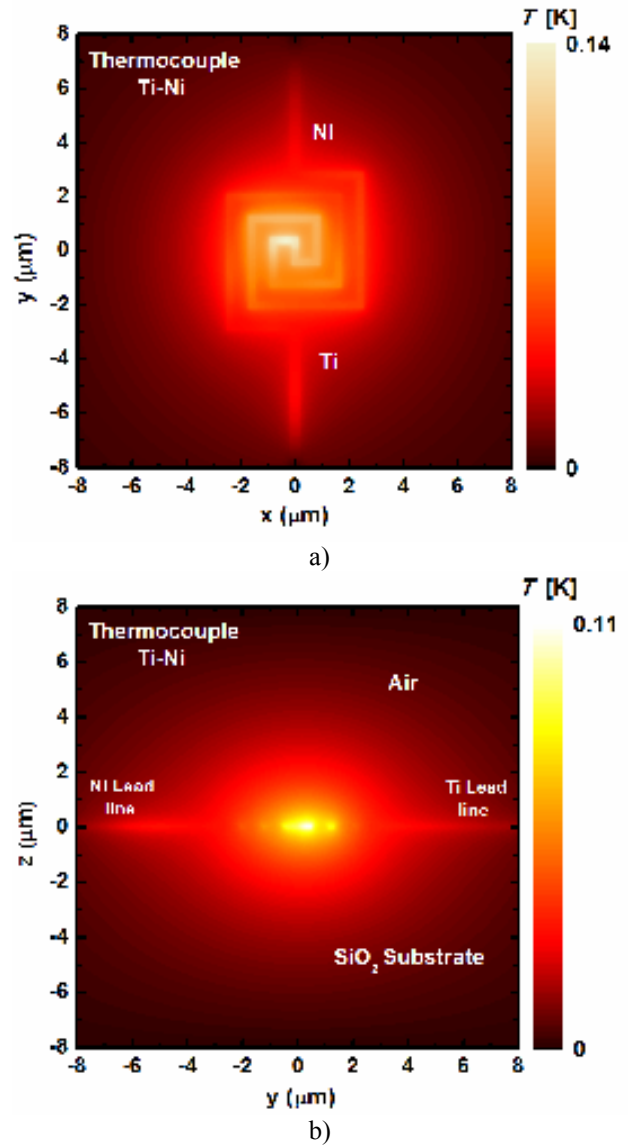

Fig. 2. Electro-thermal simulation showing the temperature distribution in a thermocouple-based spiral nanoantenna composed of a Ti-Ni bimetallic junction a) plane parallel to the nanoantenna and b) plane transversal to the nanoantenna.

By the other hand, the transversal trace shown in Fig. 2 indicates that a considerable quantity of the optical-induced heat is dissipated toward the substrate and environment, reducing this way the effective temperature increase of the thermocouple junction.

Using the temperature increase of the nano-antenna (~130mK) the Seebeck voltage all along the structure was estimated to be 2.9μV and the responsivity 20mV/W. This value is higher than several reported responsivity values for the counterpart antenna-coupled high-speed rectifiers.

## IV. Conclusions

The response of thermoelectric Ti-Ni broadband nano-antennas to convert the optical power of wavelengths at 10.6μm into dc power was evaluated by using thermal numerical simulations, showing that devices represent an alternative way to harvest free-propagating optical energy (by exploiting the optical energy naturally dissipated as heat). The response of devices can be increased by reducing the effective thermal conductivity of the substrate. This can be achieved by suspending the device on air above its substrate. Moreover, engineering of large phase-arrays of nano-antennas acting as series thermocouples arrays can be implemented to increase responsivity of devices.

## Acknowledgments

This work has been partially supported by the project "Centro Mexicano de Innovación en Energía Solar" from Fondo Sectorial CONACYT-Secretaría de Energía-Sustentabilidad Energética. The authors would like also to acknowledge support from CONACYT through postdoctoral grant CV-45809; to FONDECYT under project 3120059; and to La Region Lorraine.